# Single Crystal Growth of the New Pressure-induced-superconductor CrAs via Chemical Vapor Transport

Xiangde Zhu*, Langsheng Ling, Yuyan Han, Junmin Xu, Yongjian Wang, Hongwei Zhang, Changjin Zhang, Li Pi, Yuheng Zhang

*High Magnetic Field Laboratory, Chinese Academy of Sciences, Hefei 230031, P. R. China*

**Abstract**

Mono-arsenide CrAs, endures a helical anti-ferromagnetic order transition at ∼ 265 K under ambient pressure. Recently, pressure-induced-superconductivity was discovered vicinity to the helical anti-ferromagnetic order in CrAs[Wei Wu *et al.*, Nature Communications 5, 5508 (2014).]. However, the size of crystal grown via tin flux method is as small as 1 mm in longest dimension. In this work, we report the single crystal growth of CrAs with size of $1 \times 5 \times 1$ mm$^3$ via chemical vapor transport method and its physical properties.

**1. Introduction**

Since the discovery of superconductivity in rare-earth iron pnictides[1, 1], intensive research efforts have been exerted to explore new superconductor in transition metal oxide pnictides, pnictides and chalcogenides[3, 4, 5, 7], which leads to booming new superconductors in the past several years. Superconductivity is often found to be in the vicinity of the anti-ferromagnetic(AFM) in newly discovered iron based superconductors. Therefore, it is now widely accepted that the origin of superconductivity is related to the magnetism in iron based superconductors as well as the high temperature cuprates superconductors.

---

*Corresponding authors at: High Magnetic Field Laboratory, Chinese Academy of Sciences, Hefei 230031, China. TEL: +86-551-65595619; FAX: +86-551-65591149; Email: xdzhu@hmfl.ac.cn.

CrAs, a mono-arsnide, adopts a structure of the MnP type, which is tetragonal (space group Pnma, 62) with lattice parameters $a$ = 5.651 Å, $b$ = 3.465 Å, $c$ = 6.209 Å. The crystal structure viewed along $b$-axis and $c$-axis is shown in Fig.1a and Fig. 1b, respectively. Obviously, the crystal structure of CrAs can be regarded as stacking the unit cells along $b$-axis. The Cr atoms are six-fold coordinated by the As atoms, and lie in a zig-zag line along $a$-axis and $c$-axis. For decades ago, neutron diffraction studies demonstrate that CrAs endures a special double helical AFM order transition with a spiral propagation vector of 0.353·$2\pi c$[8]. Its Néel temperature($T_N$) was reported as 260-270K[8], 265K[9], and 240K[10]. The transition is first order, showing a hysteresis between 260-270K[8]. Below the transition, the $b$-axis lattice parameter $b$ increases by about 4%, while the $a$-axis lattice parameter ($a$) and $c$-axis lattice parameter ($c$) decreases by about 0.3% and 0.5%, respectively[11]. Just recently, superconductivity was observed in CrAs under pressure ($P$) when its helical AFM transition was suppressed, by two different groups Wei Wu *et al.*[12] and Hisashi Kotegawa *et al.*[13]. The maximum superconducting transition temperature($T_C$) is about 2.2 K at critical pressure ($P_c$) for both groups. The $P$ dependent phase diagram is reminiscent of doped iron based superconductors. Moreover, nuclear quadrupole resonance results indicate that it is a unconventional superconductor[14].

The single crystals were grown via tin flux method by both two groups[13, 15]. However, the single crystals were as small as about 0.15 × 1 × 0.15 mm$^3$, which hinders some further physics measurements. Another mono-arsnide, FeAs has been grown successfully via chemical vapor transport (CVT) method[16]. Therefore, it is worthy to try to grow CrAs single crystal via CVT method. In this work, we report the single crystal growth of CrAs with typical size of 1 × 5 × 1 mm$^3$ via chemical vapor transport method and its physical properties.

## 2. Material and methods

**Synthesis** The single crystal of CrAs was successfully grown via chemical vapor transport method with iodine as the transport agent. Stoichiometric Cr pieces(99.9%, Alfa), As pieces(99.99%, Aladin), and 50 mg iodine(99.99%, Aladin) were mixed and sealed in an evacuated quartz tube with diameter of 17 mm, and length of ∼ 20 cm. Then the tube was put in an two zone horizonal furnace. First, the temperature of both zone was slowly heated to 800 °C, and kept for four days. Second, the source zone was slowly increased to 900 °C for one week before furnace cooling. Single crystals of CrAs in shape of a helical needle can be found with the size of ∼ $1 \times 5 \times 1$ mm$^3$, which is shown in Fig. 1b. This size is significantly larger than that of the samples grown via Sn flux method [13, 15]. The as-grown crystal is covered by some black powder, which can be erased and cleaned. The cleaned single crystal is stable in air, showing shiny silver-like faces. Arsenic single crystals with hexagonal plate can also been found as a by-product. The as-grown CrAs crystals can be easily cleaved along *b*-axis.

**X-ray Diffraction and Elemental Analysis** Several single crystals were crushed and ground for powder X-ray measurement on the Rigaku-TTR3 x-ray diffractometer using high intensity graphite monochromatized Cu K$_\alpha$ radiation. Elemental analysis was performed by using energy-dispersive X-ray spectroscopy (EDS) on an FEI Helios Nanolab 600i. The EDS results gives a Cr:As ratio of 49.82:50.18, which indicates the good stoichiometry of the as-grown sample.

**Transport Properties and Magnetism** Electrical transport measurement was carried out on a Quantum Design Physical Property Measurement System(PPMS), with the current applied along *b*-axis. Data were collected over a temperature range of 2 - 300 K. Temperature dependent resistivity measurement were performed using a four-probe configuration. Gold wires were attached on a polished sample with the electrical contacts made of silver paint. The magnetization was measured by a magnetic property measurement system (Quantum Design MPMS 7T-XL) with a superconductive quantum interference device (SQUID).

## 3. Results and Discussions

### x-ray diffraction Pattern

Figure 1 shows the powder x-ray pattern of the CrAs. The peak positions are labeled and consistent with the calculated results from other literature. Some minor peaks can be observed, which should be the black oxides covered on the surface.

### Resistivity

Figure 2 shows the temperature dependence of *b*-axis resistivity($\rho_b$) with cooling and warming processes. The curves show metallic behavior with abrupt hysteresis transitions at $T$ = 247 K and $T$ = 257 K for cooling and warming processes, respectively. The residual resistivity ratio (RRR, here we use $\rho(250K)/\rho(5K)$) is $\sim$ 9, which is smaller than the crystal grown via tin flux method [15]. The transition temperature 247-257K is slight smaller than 260-270K for the crystal grown via tin flux method. For the first cooling process, the $\rho_b$ increases abruptly at $T$ = 246 K with $\sim$ 80% ($\Delta\rho_b/\rho_b$). For the second warming process, the $\rho_b$ increases with $\sim$ 68%, rather than return to the previous value for W. Wu *et al.*'s sample[12]. It is interesting that $\rho_b$ of W. Wu *et al.*'s sample decreases with decreasing temperature during the transition[12], which is different from our samples and H. Kotegawa *et al.*'s samples[13]. The *b*-axis lattice parameter expands about 4% after the helical AFM transition[11], which leads to cracks in the crystal. The cracks will result in the decrease of cross-section of electrical transport, which is the origin of resistivity jump at the AFM transition. Although the RRR and transition temperature is different from that of the tin-flux grown sample, the resistivity jump ratio 80% of our sample is significantly larger than $\sim$ 35% for the crystal grown via tin flux method [13, 15]. This indicates our sample should have larger change on cell volume. In $CrAs_{1-x}P_x$[11], it was found that the $T_N$ is related to the *b*-axis lattice parameters. The different single crystal growth condition, growth temperature and cooling procedure, may lead to different stress in as-grown crystal. The slight difference on stress and *b*-axis lattice parameter should be the origin of these difference mentioned above.

As shown in the inset of Fig. 2, the $\rho_b$ - $T^2$ curve of CrAs shows linear relation below T < 20 K, which can be well described as $\rho = \rho_0 + AT^2$, where $\rho_0$ is the residual resistivity, and $A$ is a constant. This indicates that CrAs has a Fermi-liquid ground state, in which the electron-electron interactions play the major role in the scattering mechanism. Since the cracks occur in the crystal below $T_N$, the $A$ is not intrinsic. We can estimate the intrinsic $A$ by a correction factor of $\rho_b(247K)/\rho_b(244K) \sim 0.57$, since the cracks only reduce the cross-section of the electrical transport. From the obtained fitting values of $A = 0.00241 \mu\Omega cm K^{-2}$, real $A$ is estimated as $0.00137 \mu\Omega cm K^{-2}$. With the electronic specific heat coefficient $\gamma$ = 9.1 from ref[15], the Kadowaki-Woods (KW) ratio $A/\gamma^2$ can be evaluated as $1.65\times10^{-5}$ $\mu\Omega$ cm $K^2$ $mJ^{-2}$, which is slightly larger than the universal line $a_0 = 1\times10^{-5}$ of correlated electron
systems. This demonstrates that CrAs is a correlated electron system.

**Magnetization** Figure 3 shows the temperature dependence of magnetization ($M - T$) measured by zero-field cooling process and field-cooling process with applied field ($H$) perpendicular and parallel to *b*-axis, respec-tively. Obviously, a transition with hysteresis between 250K and 258K can be observed on the $M-T$ curves, which is consistent to the resistivity results. Interestingly, the AFM transition leads to the decrease of magnetization $M$ for $H \parallel b$, while the AFM transition leads to the increase of $M$ for $H \perp b$.

Below the helical AFM transition, the magnetization decrease with decreasing temperature, and then reach the minimum around 30 K for both $H \perp b$ and $H \perp b$. These results are consistent with that reported on the single crystal grown via Sn flux method[15]. The inset of Fig. 3 shows the magnetic hysteresis of CrAs single crystal measured at 2 K for both $H \perp b$ and $H \perp b$. The $M-H$ curves show linear field dependent relation both for $H \parallel b$ and $H \parallel b$. Generally, the magnetization of our sample shows similar
behavior to that of the crystal grown via tin flux method.

## 4. Conclusions

In summary, large single crystals of CrAs with dimensions of $\sim 1 \times 5 \times 1$ mm$^3$ were successfully grown via the chemical-vapor-transport method. The phase was identified by powder X-ray. The electrical resistivity and magnetization were measured. The electrical resistivity results indicate that CrAs grown via chemical-vapor transport method has some different behavior from the crystal grown via tin flux method, while magnetization results indicate that they have similar behavior except the difference on Néel temperature.

## 5. Acknowledgements

We are very grateful to and Mrs Hui Han and Dr. Lei Zhang for the help of Powder X-ray experiments. Work at High magnetic field lab (Hefei) was supported by National Natural Science Foundation of China (Grants No. 11204312, 11474289).

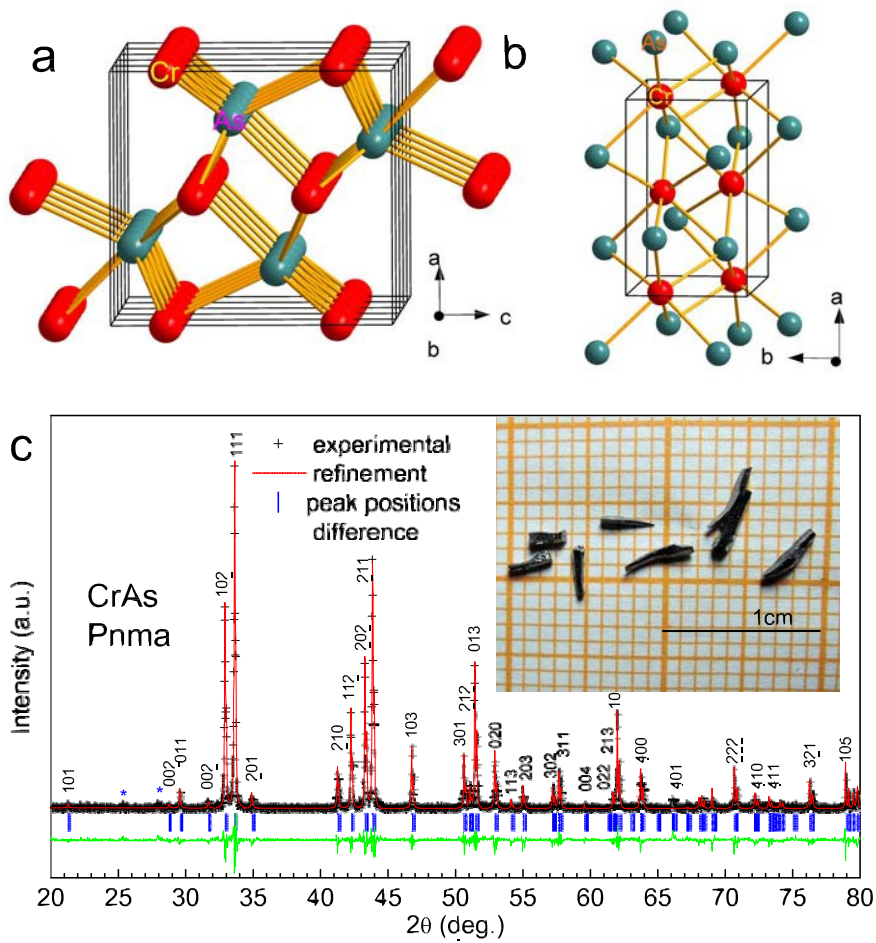

Figure 1: **a**. The crystal structure viewed from *b*-axis for CrAs. **b**. The crystal structure viewed from *c*-axis for CrAs. **c**. The experimental and refined powder x-ray patterns of CrAs. The calculated peak positions are marked as blue bars. The green line below represents the difference between the experimental and refined data. The inset of *c* shows the photograph of as-grown CrAs single crystal.

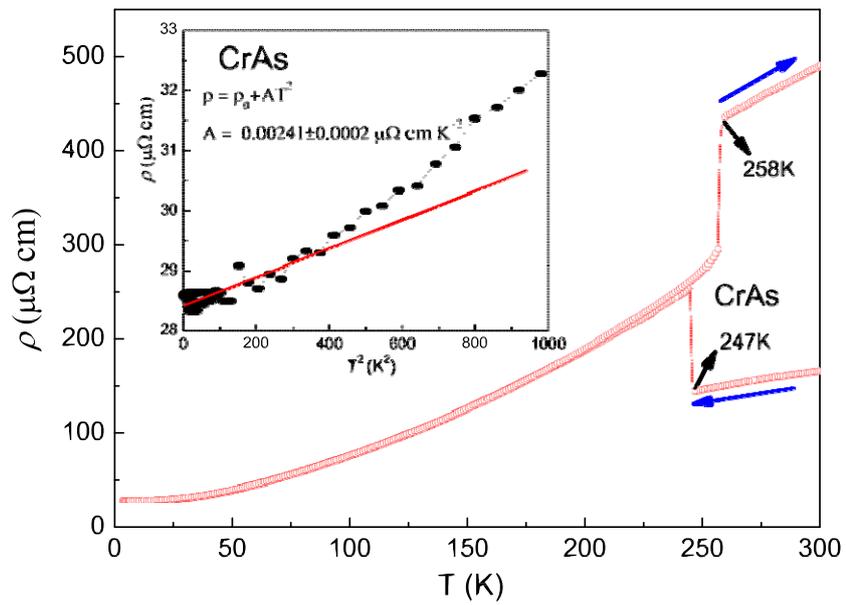

Figure 2: The temperature dependence of resistivity measured with cooling and warming process, which is marked as arrows. The inset shows the $T^2$ dependence of resistivity. The solid line represents the linear fitting results.

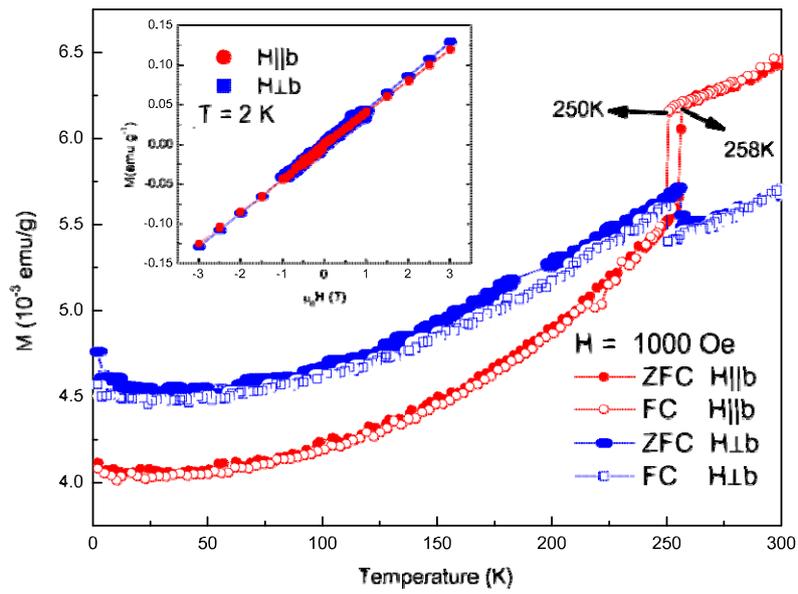

Figure 3: The temperature dependence of magnetization measured by zero-field cooling (ZFC, solid symbol) and field-cooling (FC, open symbol) process with applied field ($H$) perpendicular (rectangle symbol) and parallel (circle symbol) to $b$-axis. The $H$ is 1000 Oe in all measurements. The inset shows the magnetic hysteresis of CrAs single crystal measured at 2 K for both $H \parallel b$ and $H \perp b$.